Growth of narrow-neck, epitaxial and nearly spherical Ge nanoislands on air-exposed Si(111)-(7×7) surfaces

K. Bhattacharjee[1,#], A. Roy[2], S. Roy[2], J. Ghatak[1,$], S. Mathew[1,&], P. V. Satyam[1] and B. N. Dev[2,*]

[1]Institute of Physics, Sachivalaya Marg, Bhubaneswar 751005, India

[2]Department of Materials Science, Indian Association for the Cultivation of Science, Jadavpur, Kolkata 700032, India

---

[*] Corresponding Author. E-mail: msbnd@iacs.res.in. Phone: +91 33 2473 4971 (Ext. 200). Fax: +91 33 2473 2805

[#] Present address: Max-Planck Institute for Microstructure Physics, Weinberg 2, D-06120 Halle, Germany

[$] Present Address: Dept. of Material Science and Engineering, National Cheng Kung University, 1 University Road, Tainan, Taiwan

[&] Present address: National University of Singapore, Singapore

# Abstract

Growth of narrow-neck, epitaxial as well as non-epitaxial and nearly spherical Ge islands on air-exposed Si(111)-(7×7) surfaces has been investigated by *in-situ* scanning tunnelling microscopy (STM) and *ex-situ* high resolution cross-sectional transmission electron microscopy (HRXTEM). A thin oxide is formed on Si(111)-(7×7) surfaces via air exposure. Ge islands are grown on this oxide. STM measurements reveal the growth of very small (~2 nm diameter) Ge islands with a high number density of about $1.8 \times 10^{12}$ cm$^{-2}$. The island size has been found to depend on the amount of deposited Ge as well as the substrate temperature during Ge deposition. HRXTEM micrographs reveal that the islands are nearly spherical in shape, making narrow-neck contact with the substrate surface. At 520°C growth temperature both epitaxial and non-epitaxial islands grow. However, at 550°C, Ge islands predominantly grow epitaxially by a narrow-contact with Si via voids in the oxide layer. Growth of vertically elongated Ge islands is also observed in HRXTEM measurements with a very small diameter-to-height aspect ratio (~0.5−1), a hitherto unreported feature of epitaxial Ge growth on Si surfaces. In addition, stacking fault and faceting are observed in islands as small as 5 nm diameter. Ge islands, not even in contact with the Si substrate, appear to be in epitaxial alignment with the Si substrate. The island size distribution is essentially monomodal. As the contact area of Ge islands with Si through the voids in the oxide layer can be controlled via growth temperature, the results indicate that tunability of the potential barrier at the interface and consequently the tunability of electronic levels and optical properties can be achieved by the control of growth temperature.

# 1. INTRODUCTION

Nanoscale materials of indirect band-gap semiconductors, such as Ge nanodots, are widely investigated in order to study quantum confinement and their promising applications, especially in optoelectronics [1,2]. For practical implementation in devices, the nanodots should be very small and uniform in size with a high number density. Quantum confinement effect, arising due to the confined charge carriers inside these nanostructures, becomes strong with decreasing size of the quantum dots. Smaller the size, more prominent is the discreteness of the energy levels of the confined electrons which may lead to an increased band gap relative to the bulk, thus, showing a strong confinement effect even at room temperature (RT) [3–8]. Various approaches, starting from lithographic technique to self-organized growth phenomena have been considered so far for the spatial ordering and uniform size distribution of these nanostructures. As far as applications are concerned, in addition to optoelectronic and electronic applications, there are new possibilities of magnetic applications as ferromagnetism has recently been reported in Ge nanoparticles [9,10].

In the case of strain-driven self-organized growth of Ge on Si surfaces, Ge follows the Stranski-Krastanov growth mode with the formation of 3-D islands on top of a wetting layer. However, this conventional self-organized growth produces relatively large Ge islands [11–14]. Ge growth on a pre-grown ultrathin $SiO_2$ film on atomically clean Si(111)-(7×7) surfaces gives rise to ultra small (<10 nm) closely packed self-organized epitaxial Ge islands [6,7,15–17]. The ultrathin $SiO_2$ film saturates dangling bonds on Si(111)-(7×7) surfaces, lowers the surface free energy and modifies the kinetics and energetics of island nucleation on surfaces. Under appropriate conditions, the $SiO_2$ film reacts with deposited Ge and voids of diameter less than ~ 1 nm are formed in the $SiO_2$ film [18]. The voids work as nucleation centres, and further deposition of Ge leads to the growth of epitaxial Ge nanodots, connected through the voids to the Si substrate. Thus, adequate conditions for the growth of very small size Ge nanostructures with a proper control of the size have been possible to achieve.

The confining potential barrier height of a nanodot is regulated by interface condition [17]. Proper tuning of various parameters like growth temperature and deposition rate can help in modifying the interface conditions of the Ge quantum dots as well. Nakayama et al. [17] have shown that the confining potential for the epitaxial nanodots, in narrow-contact with the Si substrate surface, was significantly reduced compared to that of the nonepitaxial nanodots in contact with a thin oxide on Si. This provides a new way to tune quantized energy levels of Ge nanodots not only by their size but also by interface condition. A reduced potential barrier of epitaxial nanodots can confine the carriers even though they are connected to the substrate [17]. Growth of Ge nanoislands by other means can also be used to tune this confining potential barrier. For example, Ge islands grown on a thin polymer layer on Si [19] would give a high potential barrier. Variation in the confining potential barrier would affect the carrier exchange at the dot-substrate interface, an understanding of which is important for applications of these nanodots for optoelectronic devices. Distinct electronic transport behaviours (I-V curve) for coherent and incoherent Ge quantum dots, depending on the confining potential, were reported by Chung et al. [20].

For Ge nanoisland growth on a thin oxide on Si, in earlier experiments the oxide layer was grown under ultrahigh vacuum (UHV) condition [3, 4, 6, 7, 16–18]. Several earlier experiments performed under atmospheric or non-UHV conditions achieved results comparable to those obtained under UHV conditions on atomically clean Si(111)-(7×7) surfaces [21–24]. It would be interesting to investigate the growth behavior of Ge nanoislands when a non-UHV step is introduced. In the present work we investigate the growth of Ge nanoislands on a thin oxide on Si that is prepared by exposing clean Si(111)-(7×7) surfaces to air.

Here we present our results on the growth of nearly spherical, small and epitaxial Ge nanodots on Si by a method of interface modification by exposing clean Si(111)-(7×7) reconstructed surface to air, giving rise to the formation of a thin (~2 nm) $SiO_x$ layer, prior to Ge deposition. Results of *in-situ* scanning tunnelling microscopy (STM) and *ex-situ* cross sectional transmission electron microscopy (XTEM) investigations are presented. We observe growth of both coherent and incoherent but well-aligned Ge islands. Coherent islands grow through the voids in the $SiO_x$ layer. For the growth of Ge nanodots on air-exposed Si(111)-(7×7) surfaces we even observe epitaxial islands with aspect ratio (diameter-to-height) of <1, some as small as 0.5, an hitherto unreported feature. Though, STM provides the direct mapping of the surface morphology, this technique is not suitable when it is essential to study the correct shape of the islands, the epitaxial quality and the correct base diameter of the nanodots grown on the surface. It is often very likely, that the base diameter provided by STM is not correct. If the contact regions of the nanodots at their base are narrower than the diameter of the dots, this narrow-neck feature will only be seen by techniques like XTEM. Therefore, high resolution XTEM (HRXTEM) studies have been used here extensively to investigate, in addition to crystallographic orientation, the island shape and their interface conditions, which are not revealed by usual STM studies. All these aspects of Ge nanodots grown on Si(111)-(7×7) surfaces, modified by air-exposure, will be addressed here.

## 2. EXPERIMENT

Ge deposition was carried out in a custom-built ultrahigh vacuum (UHV) chamber where the base pressure inside the growth chamber was of the order of low $10^{-10}$ mbar to high $10^{-11}$ mbar. A commercial UHV variable temperature scanning tunnelling microscope (VTSTM, Omicron Nanotechnology, Germany) is connected to the growth chamber for *in-situ* characterizations of the samples. This system was described elsewhere [14]. P-doped n-Si(111) samples (oriented within ±0.5°) with a resistivity of 5-20 Ω-cm was loaded in the UHV chamber. Atomically clean, reconstructed Si(111)-(7×7) surfaces were prepared by usual heating and flashing procedure. The samples were first degassed at about 600°C for 12-14 hours followed by a prolonged flashing at ~1200°C for 5−6 minutes. The substrate was then cooled down to room temperature (RT) and Si(111)-(7×7) surface reconstruction was observed by *in-situ* STM. Step-terrace structure on the surface with the formation of defect lines on each terrace was observed. High resolution STM images show atomically clean Si(111)-(7×7) surfaces within domains separated by defect lines and step edges. The samples were then exposed to air for 3-4 hours by bringing them out via the load lock chamber. This is expected to form a thin $SiO_x$ film on the Si(111)-(7×7) surface. Following this, the sample was again introduced into the UHV growth chamber and degassing of the sample was carried out at around 500°C for 2 hours. This procedure was carried out to remove adsorbed gases from the sample surface due to air-exposure without removing the thin oxide film formed on the Si(111)-(7×7) surface. After cooling down the samples to RT, the surfaces were investigated by STM. Different samples were prepared for Ge coverages of 1.6, 2.7, 3.2, 4.8 and 6.4 bilayers (BL) (keeping all other growth parameters same) on $SiO_x$-covered Si(111)-(7×7) surfaces from a PBN (pyrolytic boron nitride) crucible at a substrate temperature of 520°C. (1 BL of Ge atoms in the [111] direction is equivalent to $1.44×10^{15}$ atoms/cm$^2$. In addition, one sample was prepared by depositing a 2.7 BL Ge layer at a substrate temperature of 550°C. The growth rate of Ge for all the coverages was kept at 0.02 Å/s and the pressure inside the UHV growth chamber during deposition was $1.7×10^{-10}$ mbar. The amount of Ge deposition during growth was measured by a quartz microbalance which has been pre-calibrated by Rutherford backscattering spectrometry (RBS) measurements on Ge films grown at different growth rates and for various amounts of Ge deposition. The post-growth investigations of the samples were carried out by *in-situ* STM and *ex-situ* HRXTEM.

## 3. RESULTS AND DISCUSSIONS

### 3.1. STM Results

#### *3.1.1. Clean and air-exposed Si(111)-(7×7) surfaces*

A STM image [Fig. 1(a)] of a clean Si(111)-(7×7) surface reveals the presence of monatomic steps and terraces with domain boundaries. High resolution STM images show atomically clean Si(111)-(7×7) surface reconstruction. A STM image of a domain boundary with Si(111)-(7×7) atomic reconstruction on both sides of it is shown in Fig. 1(b). The width of the domain boundary regions was found to vary depending on the duration of flashing. In our case, it is typically from 5−20 nm in width. The bright contrast at the domain boundaries may be because of accumulation of some contaminants. There are many reports available which confirm the formation of various kinds of defects on Si(111)-(7×7) surfaces [25–31]. Demuth et. al. [28] observed extended defect line structures, similar to our case, which these authors attributed to surface grain boundary, either due to short annealing period or because of some contaminants on the surface. Since, the domain boundaries on Si(111)-(7×7) surfaces are irregular [31], proper reasons behind the formation of such defects are difficult to ascertain. These boundary structures are mostly due to the strong interaction between dimer and adatom, the difference between faulted and unfaulted halves and sometimes due to the metastable subunit cells like (5×5) [27]. In our case, we have observed that a prolonged flashing time (5 minutes or more) gives rise to domain boundaries on Si(111)-(7×7) surfaces.

When a Si(111)-(7×7) surface [Fig. 1(a)] is exposed to air a thin oxide layer is formed. Surface morphology of the thin $SiO_x$ film grown on Si(111)-(7×7) surfaces is shown in figure 2. STM investigations reveal that a uniform $SiO_x$ film has formed on the surface, keeping the step-terrace structure of the clean Si surface intact. The step height, as seen from the STM height profiles (not shown) after the formation of a $SiO_x$ film, remains almost the same as that of the clean (7×7) surfaces [Fig. 1(a)]. Oxide growth appears to be uneven on the domain boundaries, unlike on the (7×7) reconstructed terraces where it appears almost uniform.

Oxidation of passivated Si surfaces in air is a slow process related to desorption rate of the passivating species and somewhat dependent on crystallographic orientation of the surface [32]. Investigations regarding the oxidation process on H-terminated Si(001) surfaces show a 0.5 nm thick oxide layer growth [33] on the surface at RT after keeping the sample for one day in air where the step-terrace structure was found to be preserved even after the oxidation. However, clean surfaces, when exposed to air, undergoes the oxidation process comparatively rapidly with a sticking coefficient of $10^{-4} - 10^{-1}$ [34]. In the present case a 3–4 hr. exposure to air was sufficient to form a thin oxide. Ge nanodots have grown on this surface upon Ge deposition.

It was shown earlier for the growth of oxide under UHV condition on Si(111)-(7×7) surfaces that the (7×7) reconstruction of the Si(111)-(7×7) surface remains on the oxide surface [35, 36]. For the oxide grown by air-exposure, we also observe a short range (7×7) order on the oxide surface. Fourier transform of the STM image of the air-exposed surface in Fig. 2 and similar other images show the spots corresponding to the 2.7 nm periodicity of the (7×7) reconstructed surface. This is shown in Fig. 3, where (1/7, 0), (0, 1/7), (1/7, 1/7) and their equivalent fractional order spots of the (7×7) reconstruction on Si(111) are marked. The distances between the spots are one-seventh the distance corresponding to the (1×1) unit cell on Si(111). This indicates that the periodicity of the underlying Si(111)-(7×7) unit cell remains on the oxide surface grown by air-exposure, although the oxide prepared by air-exposure is not likely to be as clean as the UHV-grown oxide. The thickness of this oxide layer is revealed in many XTEM images presented in part B of this section.

### *3.1.2. Ge growth on air-exposed Si(111)-(7×7) surfaces*

We have studied Ge growth at two different temperatures, 520°C and 550°C. Temperature was measured by a thermocouple at the rear side of the substrate. So the substrate surface temperature is expected to be somewhat lower. STM images of a 1.6 BL Ge film deposited at 520°C substrate temperature, are shown in Fig. 4. Densely packed Ge islands have grown on the surface, forming a 'tiling' pattern where each 'tile' is decorated with Ge islands. The 'tiles' are separated from each other by narrow boundary regions, where the growth of Ge islands is rare. Ge islands are less likely to grow on the presumably contaminated boundary region. The width of the boundary regions, as seen in the STM micrographs, is around 10–20 nm which is approximately the same as the width of the domain boundaries on the Si(111)-(7×7) surfaces (figure 1). The STM images in Fig. 4(a), (b) show a few 'tiles' and the boundary regions. A portion within a 'tile' is shown in Fig. 4(c). To study the details of the surface around the boundary regions, STM measurements were made on and around these places. A higher resolution STM image of such a 'tile' surrounded by the boundary regions is shown in Fig. 4(d). The STM micrographs reveal that very few Ge islands have grown in the boundary regions. However, very small size island-like structures are observed in these areas. Height distributions obtained from the images in Fig. 4(a), (b), (c) and (d) are shown in figure 5(a)[(i), (ii)], (b)[(i), (ii)], (c) and (d), respectively. (In each image there is 450×450 points. Heights of these 202500 points are plotted as no. of points vs. height). In Fig. 5, the peak marked as 'A' corresponds to the base layer and the peak, marked 'B' corresponds to the Ge islands. Such analysis of STM images was used earlier to determine island height distribution [37, 38]. We fitted the height distributions [Fig. 5] using Gaussian functions. The height distribution curves [Fig. 5(a)(i), (b)(i) and (c)], when fitted with two Gaussian functions, do not fit properly whereas a fit considering three Gaussian functions appears to be more appropriate [Fig. 5(a)(ii), 5(b)(ii) and 5(c)]. This implies that apart from the peak A, which appears from the base layer, and the peak B due to Ge islands, there is a small peak between A and B. This small peak appears to come from the boundary regions between the tiles. From the fitted distributions, we obtain the island heights relative to the first peak (A) position. The first set of smaller 'islands' corresponds to a height of ~0.7 nm, and the second set of islands (B), which are larger in size corresponds to a height of ~2 nm. In an STM micrograph from a small region (Fig. 4(d)) containing a large fraction of the boundary regions as well as island regions shows only one island peak. The peak A is broad and shifted to the position of the peak from the boundary region as expected. The size distribution of Ge islands is essentially monomodal. The average number density of islands, in our case, within a 'tile' is found to be about $1.8 \times 10^{12}$ cm$^{-2}$. Previous STM studies [16] of Ge growth on a UHV-grown SiO$_2$ layer (0.3 nm) on Si(111) surfaces, reported growth of Ge islands with an island height of about 2.3 nm for 2.6 BL Ge coverage, with a number density of approximately $1.8 \times 10^{12}$ cm$^{-2}$, similar to our results. These authors did not study any height distribution of the islands, only height profiles of a few individual islands were shown. We have also carried out STM studies for 2.7 and 3.2 BL Ge coverages. In the case of 2.7 BL Ge the average island size is larger, ~3 nm. For 3.2 BL Ge, islands grow further in size and overlapping islands are formed.

STM images (Fig. 4) of the 1.6 BL Ge film on $SiO_x$ reveal that the Ge islands have formed overwhelmingly on the 'defect-free' areas of the terraces. Structural defects on a substrate surface play an important role in semiconductor heteroepitaxy. Surface electronic structure changes with the occurrence of the structural defects, which results in different surface chemical properties. Therefore, the growth, electronic structure and the physical properties of the overlayer are affected by the interface defects [39–41]. The domain boundary which might appear on Si(111)-(7×7) surfaces due to various reasons, can influence the growth of the deposited material [42]. In our case, this apparently seems to be the reason behind this interesting 'domain-patterned' growth of Ge islands on modified Si(111) surfaces. The authors in ref. [16] have also not observed any evidence of the nucleation of Ge islands on the structural defect sites on $SiO_2$ films. We have also studied Ge growth on clean Si(111)-(7×7) surfaces as shown in Fig. 1(a), (b), where Ge islands preferably decorate the domain boundaries [43]. These differences can be explained by a reaction-diffusion model [44]. These works will be published elsewhere.

## 3.2. HRXTEM Results

As STM looks at the projection of the islands on the surface, very often, it can neither provide the correct shape of the islands nor can give the correct base diameter. It is very likely that the base diameter of the islands in ref. [16] is much smaller. In our HRXTEM studies, where the cross-sectional view from the side of the islands is observed, we indeed notice that the islands have usually a very narrow contact region with the substrate, much smaller compared to the base diameter observed in STM images. We observe the growth of epitaxial Ge islands connected to Si via voids in $SiO_x$ as well as non-epitaxial islands resting on the oxide layer. Use of only the STM technique is inadequate to obtain the island shape and their epitaxial or non-epitaxial nature. Details about the island-substrate contact region are important for understanding the electronic energy levels in quantum confinement [3, 7, 17, 18].

### *3.2.1. Growth of Ge at 520°C*

HRXTEM images in Fig. 6 clearly show the growth of spherical or nearly spherical, ultrasmall, epitaxial Ge islands for 1.6 BL of Ge coverage where the size of the contact area with the Si substrate is either comparable or slightly smaller than the lateral dimensions of these islands. Many islands are even found to grow nonepitaxially without any contact with the surface. These features will not be revealed by STM studies. For this coverage of Ge, we find that the islands have an aspect ratio (diameter : height) of ~1:1, much different compared to the pyramidal (10:1) and dome (5:1) islands usually observed in Ge growth on bare Si surfaces [11–13]. The islands are monomodal unlike the distinct bimodal distribution usually observed for epitaxial growth of Ge islands on bare Si surfaces [11–13].

Some HRXTEM images of 4.5 and 6.4 BL Ge films grown at 520°C are shown in Fig. 7 and Fig. 8, respectively. Ge islands for these cases are in general larger (~5 nm in dia) compared to the 1.6 BL film. For these films, we observe the growth of spherical or nearly spherical islands with very narrow contact areas with the substrate. Interestingly, some islands also show faceting [Fig. 7(b)] and stacking faults [Fig. 7(d)]. Some islands are in contact with Si (epitaxial) and some are resting on the oxide film without any contact with Si, as revealed by the XTEM micrographs. Many islands have epitaxial orientation with respect to the Si substrate, although the islands are not epitaxial as there is no contact between the Ge island and the Si substrate. The XTEM micrograph in Fig. 8(b) from a 6.4 BL Ge film shows such an island resting on the oxide film without any contact with Si. Such nonepitaxial island growth with an epitaxial orientation with the substrate was also observed for Ag growth on air-exposed Si(100)-(2×1) [45], Si(111)-(7×7) [46] and Si(110)-(5×1) [47] surfaces. This epitaxial orientation apparently arises due to the presence of short-range order on the oxide surfaces.

*3.2.2. Growth of Ge at 550°C*

XTEM micrographs of a 2.7 BL Ge film grown at 550°C substrate temperature are shown in Fig. 9. At this growth temperature, one observes very good quality epitaxial growth of Ge islands, forming narrow-neck with the Si surface. The contact area of the Ge islands with the Si substrate is, in general, much larger compared to the growth at 520°C. Islands are nearly spherical in shape and the lateral dimension of the islands is ~10–12 nm as seen in the XTEM images. The island size is found to be larger in this case compared to the size of the islands even for higher coverages of 3.2 and 4.5 BL grown at 520°C. Here also the islands have an aspect ratio (diameter : height) of ~1:1. XTEM micrographs taken from the same sample also reveal the formation of Ge islands with vertically elongated shape as shown in Fig. 10. This shape of the islands would also not be revealed by STM studies. These islands grow epitaxially, some with a very narrow contact region, and with a diameter-to-height aspect ratio < 1. Sometimes the aspect ratio of these islands is even found to be ~1:2. This kind of growth of Ge islands on Si surfaces, to our knowledge, is hitherto unreported. Ge epitaxial islands, grown on bare Si(111)-(7×7) surfaces have very different characteristics. They are pyramid- or dome-shaped with aspect ratios of about 10:1 and 5:1, respectively [12, 13]. These Ge islands, grown on bare Si surfaces, also have very large contact areas with a Ge wetting layer on the Si substrate. For the epitaxial growth of Ge islands with a narrow contact area with the Si substrate, continued deposition would prefer attachment of deposited Ge atoms with the already formed atomic strings of Ge in the island. This would tend to an elongated island growth.

The energy levels in a spherically symmetric Ge quantum dot would have degeneracies. In elongated Ge islands like those in figure 10, some of these degeneracies would be lifted and degenerate energy levels would split. Such electronic structure of individual islands can be investigated in future by scanning tunnelling spectroscopy experiments.

Even though, Ge was deposited on a Si oxide film, the HRXTEM micrographs reveal that Ge islands have grown epitaxially with a narrow-neck contact with Si via formation of voids within the oxide film. It was found that the Ge growth at an elevated temperature on a Si oxide film on Si(111)-(7×7) surfaces occurs through the following reaction [16, 48–50]:

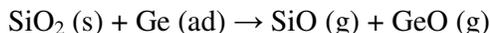

$$SiO_2 (s) + Ge (ad) \rightarrow SiO (g) + GeO (g)$$

where the evaporation of SiO and GeO molecules are significant above 500°C and 360°C respectively [16]. As a result bare Si surface becomes exposed through holes in the oxide layer created by the above reaction. This provides the necessary epitaxial growth of 3D Ge islands. This epitaxial growth of Ge islands depends on deposition rate and growth temperature [16]. It was observed that a substrate temperature of slightly above 430°C and a low deposition rate provide the adequate conditions for epitaxial growth of 3D Ge islands on an atomically clean and then oxidized Si surfaces. The voids are usually less than ~1 nm in diameter, and further deposition generates Ge nanodots connecting directly to the Si substrate through the voids [18]. Various issues related to the formation of Ge nanodots on UHV-oxidized Si(111) surfaces were addressed and discussed by Ichikawa et al. [3, 17, 18]. However, to our knowledge, there is only one report available in the literature[43] which contains XTEM images. HRXTEM images can reveal the epitaxy of the islands and the nature of the island-substrate contact region. Here, we show our HRXTEM results in details for different coverages of Ge and for two different growth temperatures. Our oxide-layer preparation is also different from that used by Ichikawa et al. and other authors [3, 17, 18]. These authors have grown the oxide layer under UHV conditions. We have grown the oxide layer by exposing clean Si(111)-(7×7) surfaces to air. Even on this presumably dirty oxide we obtain epitaxial growth of Ge islands. We also observe that among other aspects island shape and aspect ratio can be controlled by tuning the growth temperature.

## 4. SUMMARY AND CONCLUSIONS

Ge growth on air-exposed Si(111)-(7×7) surfaces has been investigated by STM and HRXTEM. Nearly spherical self-organized Ge islands with a narrow contact area with the substrate have been found to grow on air-exposed Si(111)-(7×7) surfaces. These Ge islands are usually small, in some cases as small as ~2 nm diameter depending on the growth condition. Growth at a substrate temperature of 520°C produces both epitaxial and nonepitaxial Ge islands. However, at 550°C the islands are predominantly epitaxial. Epitaxial islands grow in contact with Si through voids in the oxide layer grown by air-exposure. These voids are apparently formed via reaction of Ge adatoms with the silicon oxide. On the other hand nonepitaxial Ge islands grow where the intact oxide layer prevents growth of Ge in contact with the Si lattice. A major fraction of Ge islands grown at 550°C has an aspect ratio (diameter : height) in the range of 1:1−1:2, whereas standard self-organized growth of Ge on Si(111)-(7×7) surfaces yields islands of aspect ratios in the range, 5:1−10:1 with broad contact areas with the substrate [11–13]. In the present case, the Ge islands also have a reasonably uniform size distribution as opposed to distinct bimodal size distribution observed for self-organized growth on bare Si surfaces [11–13]. The oxide layer between the Ge islands and Si offers an electron tunnelling barrier for electron transport through these islands. A narrow contact of Ge islands with Si through the oxide layer would still offer a potential barrier, albeit smaller in comparison with the intact oxide. This indicates the tunability of the electronic levels in these Ge islands. The present work reveals the interesting result that even on the presumably dirty surface of air-exposed Si(111)-(7×7) surfaces epitaxial Ge islands can be grown. Growth of vertically elongated epitaxial Ge islands on Si, observed in the present study, to our knowledge has not been hitherto reported. Earlier studies of Ge growth were carried out on a ~0.3 nm thick UHV-grown oxide. The present work shows that epitaxial growth of Ge nanoislands occurs even on a thicker (~2 nm) oxide layer, which also need not be prepared under UHV condition. The present results also indicate that tunability of electronic energy levels and optical properties of Ge islands can be achieved by controlling the growth temperature as the contact area of the islands with the Si substrate would determine these properties.

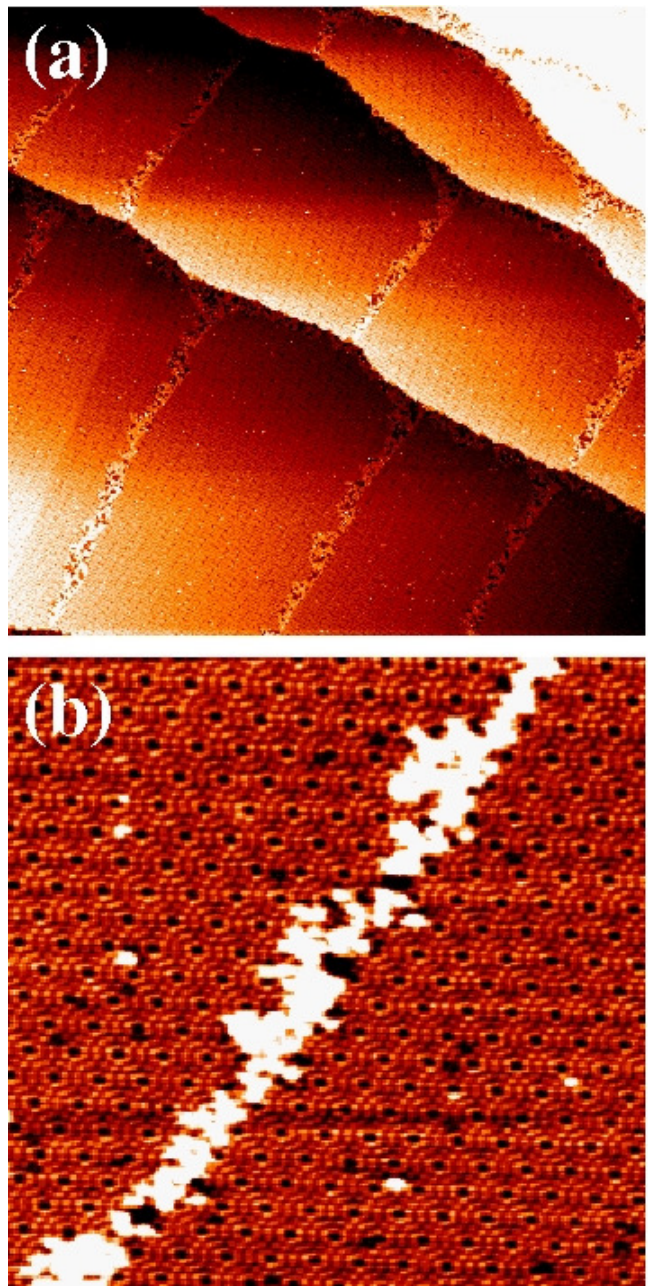

**Fig. 1.** STM images of Si(111)-(7×7) surface reconstruction, bias voltage, $V_s$ = 2.1 V, tunnelling current, I = 0.2 nA (a) steps with the formation of 10 - 20 nm wide defect lines on the surface, scan area: 800×800 nm$^2$ (b) a high resolution STM image showing a defect line with 7×7 surface reconstruction on both sides of the defect line, scan area: 40×40 nm$^2$.

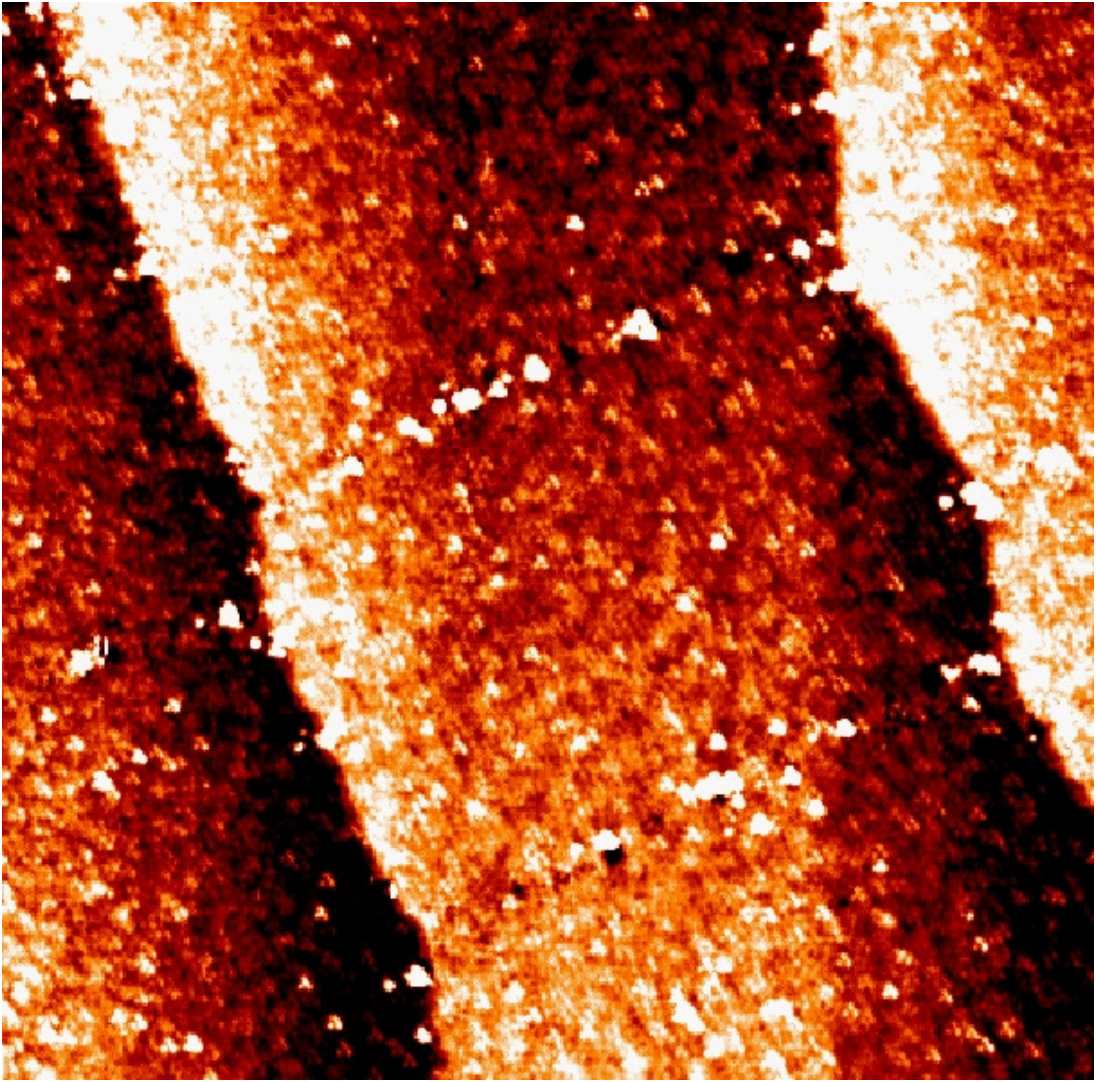

**Fig. 2.** STM image of an air-exposed Si(111)-(7×7) surface. Scan area: 525×525 nm$^2$.

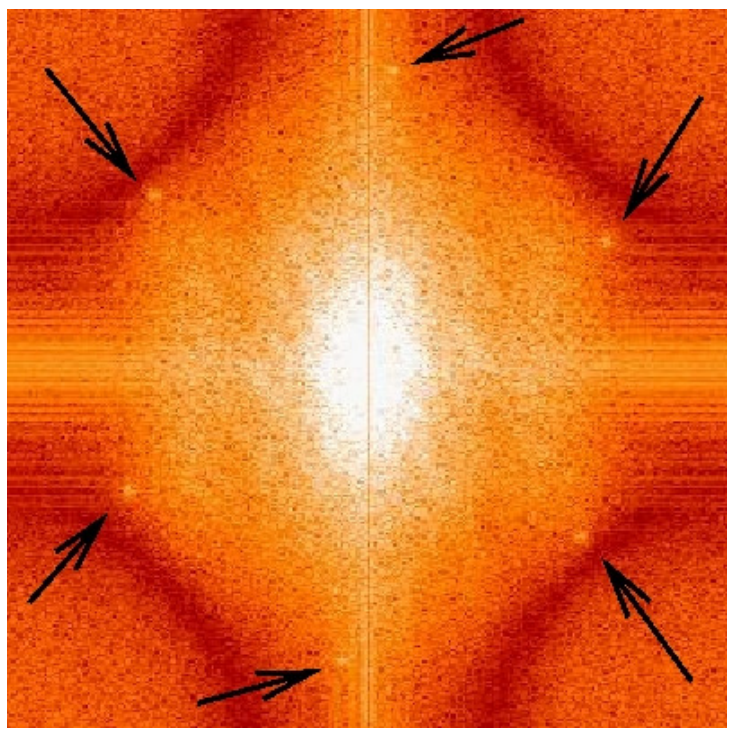

**Fig. 3.** Fourier transformed image of an air-exposed Si(111)-(7×7) surface. The spots corresponding to the 2.7 nm periodicity of the (7×7) reconstructed surface are marked.

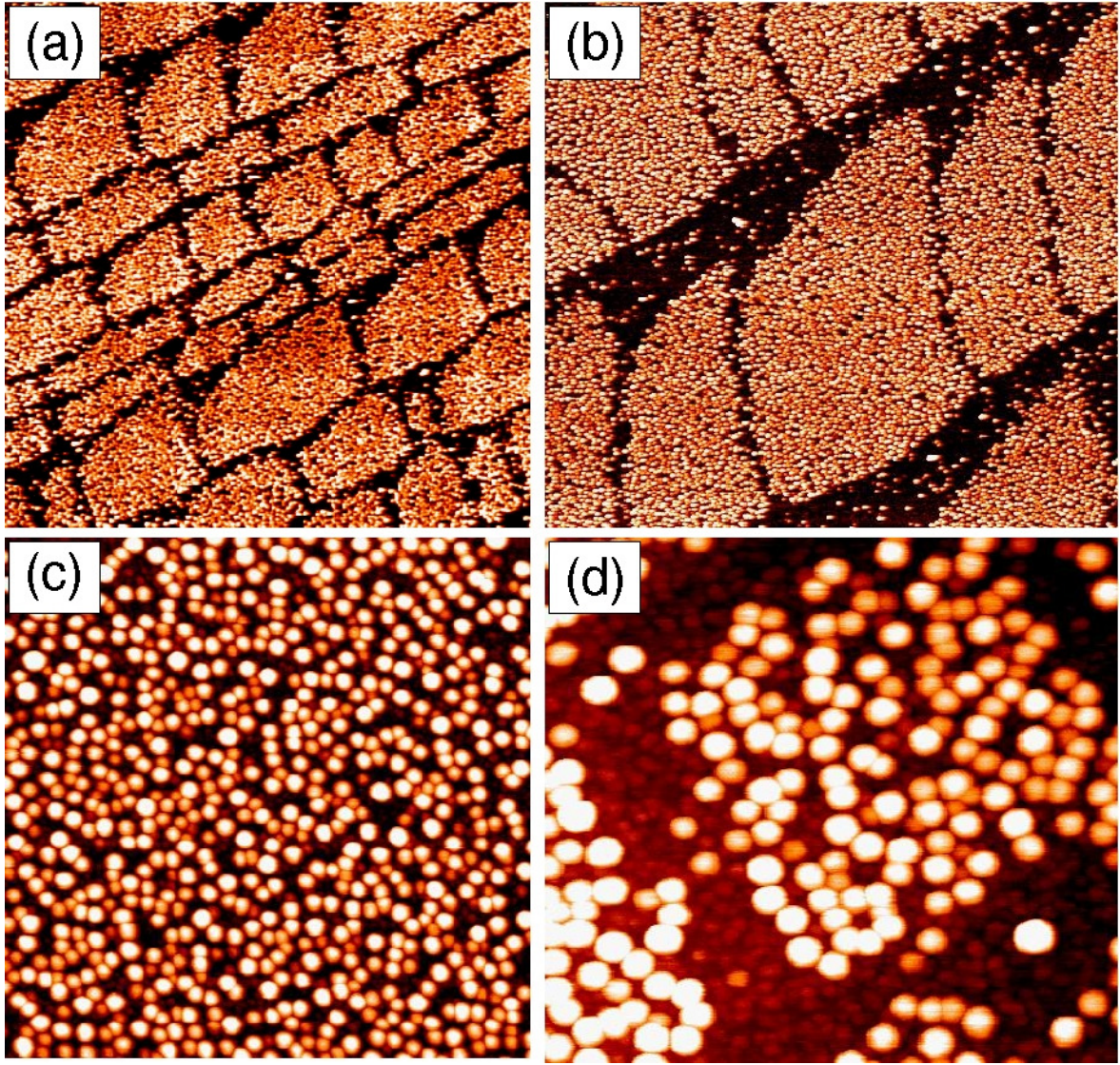

**Fig. 4.** STM images of a 1.6 BL Ge film deposited on an air-exposed Si(111)-(7×7) surface at 520°C. Sample bias voltage $V_s$ = 2.4 V, tunnelling current I = 0.18 nA. (a) 1000×1000 nm$^2$ and (b) 780×780 nm$^2$ images showing the formation of 'tiled-patterned' Ge islands (c) Ge islands within a 'tile' pattern, scan area: 200×200 nm$^2$ (d) A 'tile' pattern surrounded by defect regions where the growth of very high density, ultrasmall voids are seen in the image.

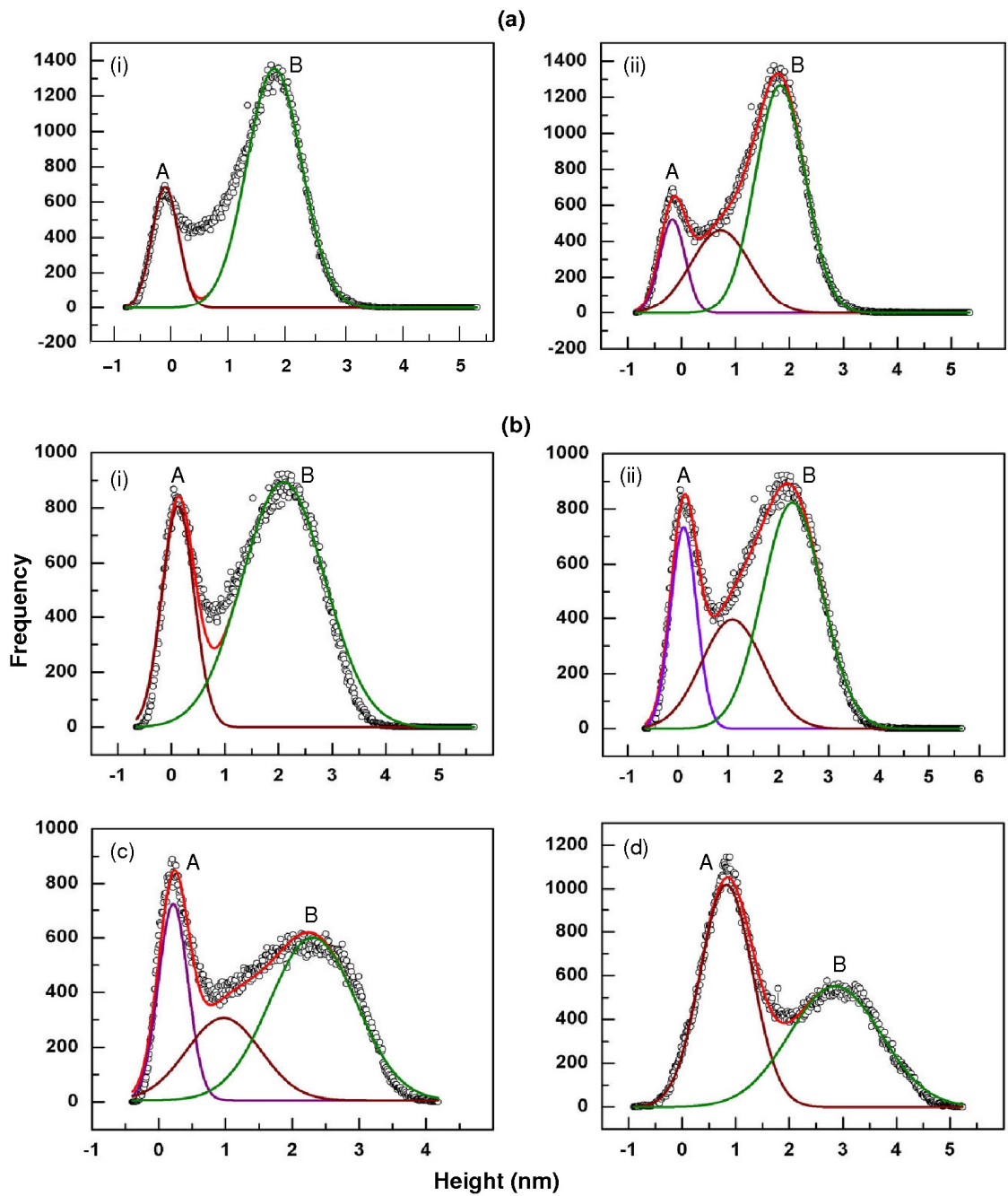

**Fig. 5.** Height distribution curves obtained from the STM images shown in figures 4(a, b, c, d). Two and three Gaussian fittings (a)[(i), (ii)] and (b)[(i), (ii)] performed respectively on the height distribution data obtained from figures 3(a) and (b) respectively. One can see that a three Gaussian fit looks more appropriate for these two images. (c) The height distribution data from figure 3(c) which could be fit only using three Gaussian distributions. (d) Height distribution data from the STM image shown in figure 3(d) fitted with two Gaussian distributions.

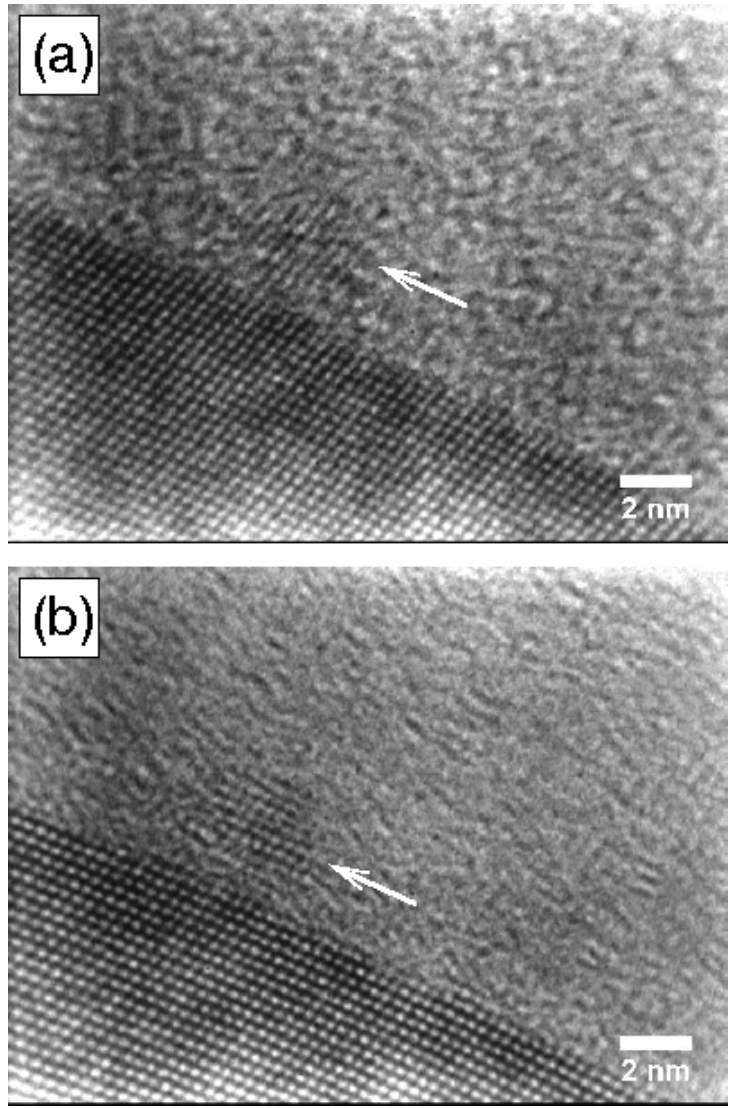

**Fig. 6.** HRXTEM micrographs obtained from a 1.6 BL Ge deposited on an air-exposed clean Si(111)-(7×7) surface at 520°C growth temperature. Ultrasmall Ge islands with narrow contact with the substrate surface are seen in the images.

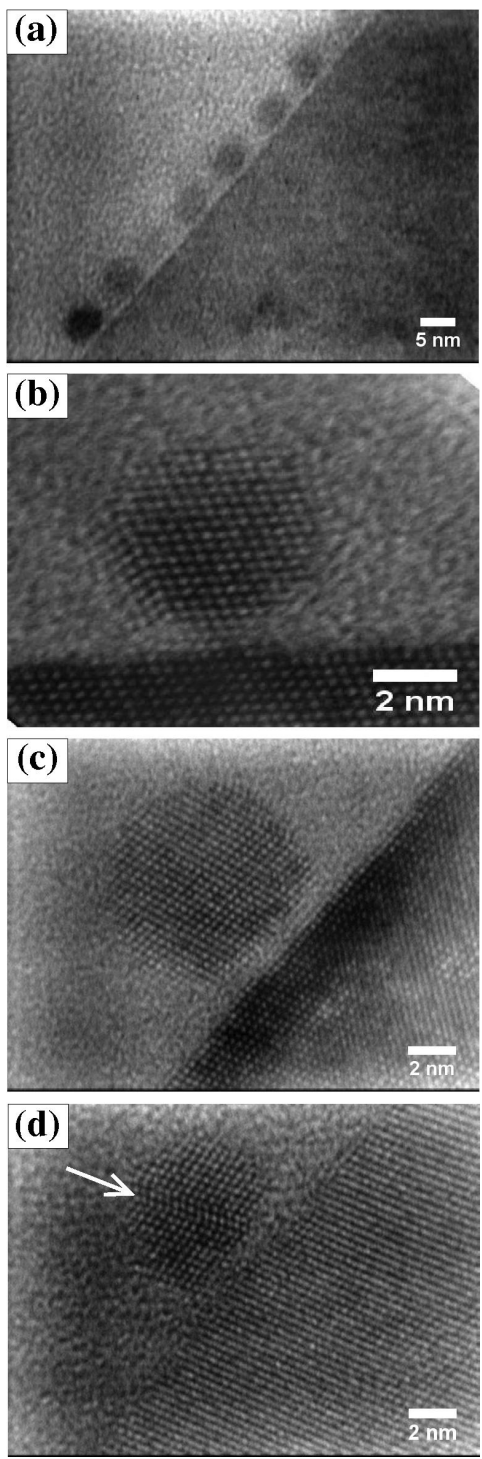

**Fig. 7.** (a) A XTEM micrograph obtained from a 4.5 BL Ge film on an air-exposed clean Si(111)-(7×7) surface grown at 520°C. HRXTEM images show (b) epitaxial (with a very narrow contact with Si) or (c) non-epitaxial Ge islands. Some images are faceted as in (b) and some have stacking faults as in (d). The oxide layer thickness as seen from (a) and (c) is ~1-2 nm.

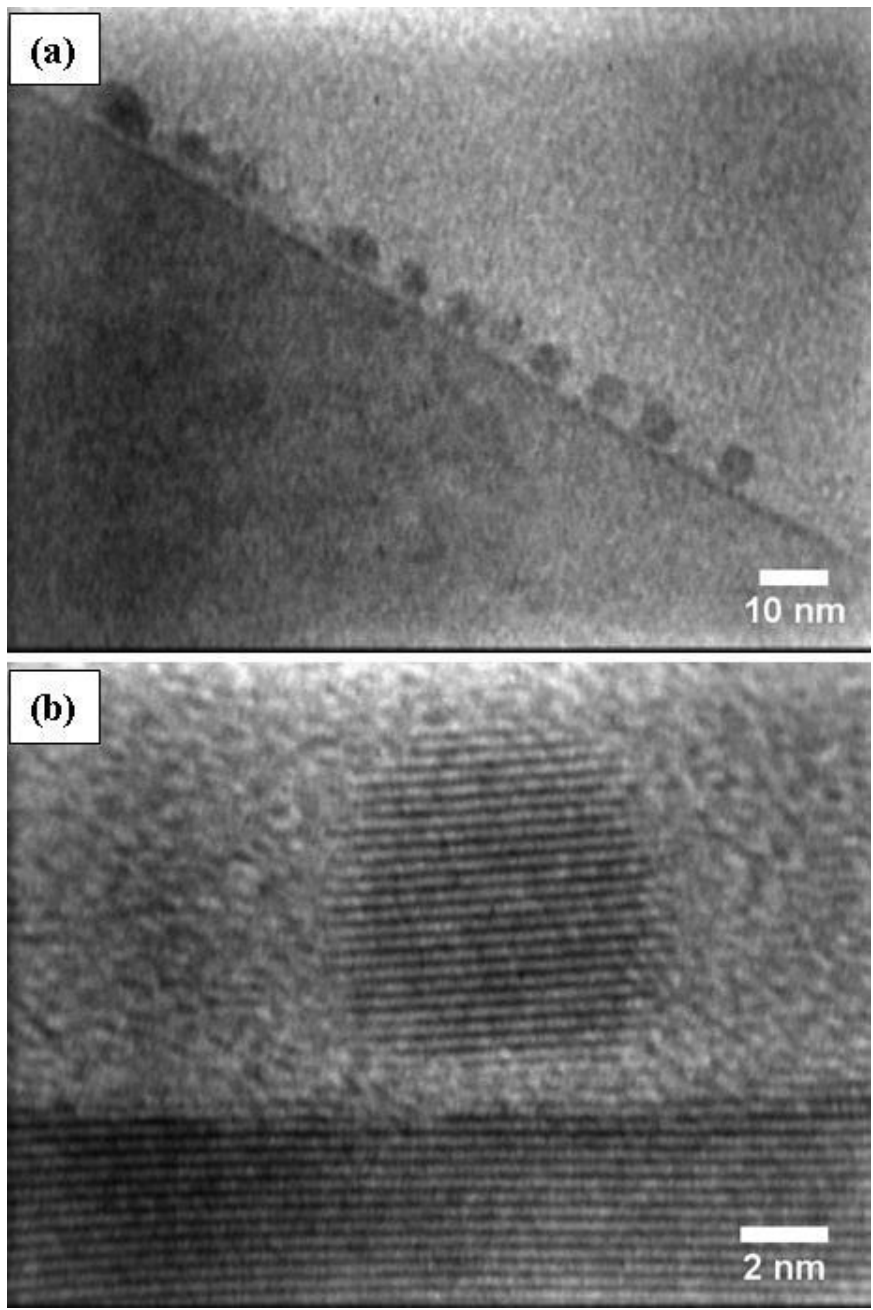

**Fig. 8.** XTEM micrographs obtained from a 6.4 BL Ge film grown at 520°C. (a) Several Ge islands are seen on the Si substrate with the oxide layer in-between. (b) HRXTEM image: the Ge island has no direct contact with the Si through the oxide. In spite of that the atomic planes in the Ge island appear to be nearly parallel to those in the Si substrate.

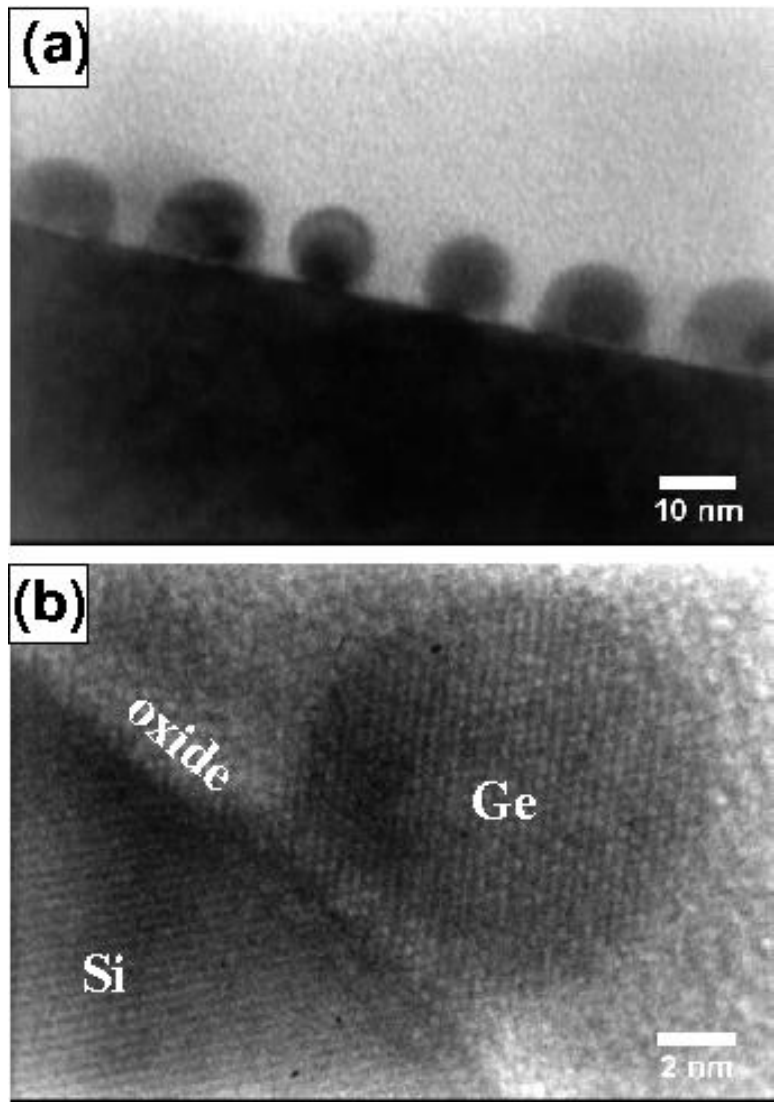

**Figure 9.** (a) A XTEM micrograph obtained from a 2.7 BL Ge film grown at 550°C on an air-exposed clean Si(111)-(7×7) surface. The islands are comparatively larger in size and have grown epitaxially forming narrow-neck with the substrate surface. These islands have an aspect ratio (diameter : height) of ~ 1:1; (b) a typical HRXTEM micrograph of a single Ge island showing the epitaxial nature.

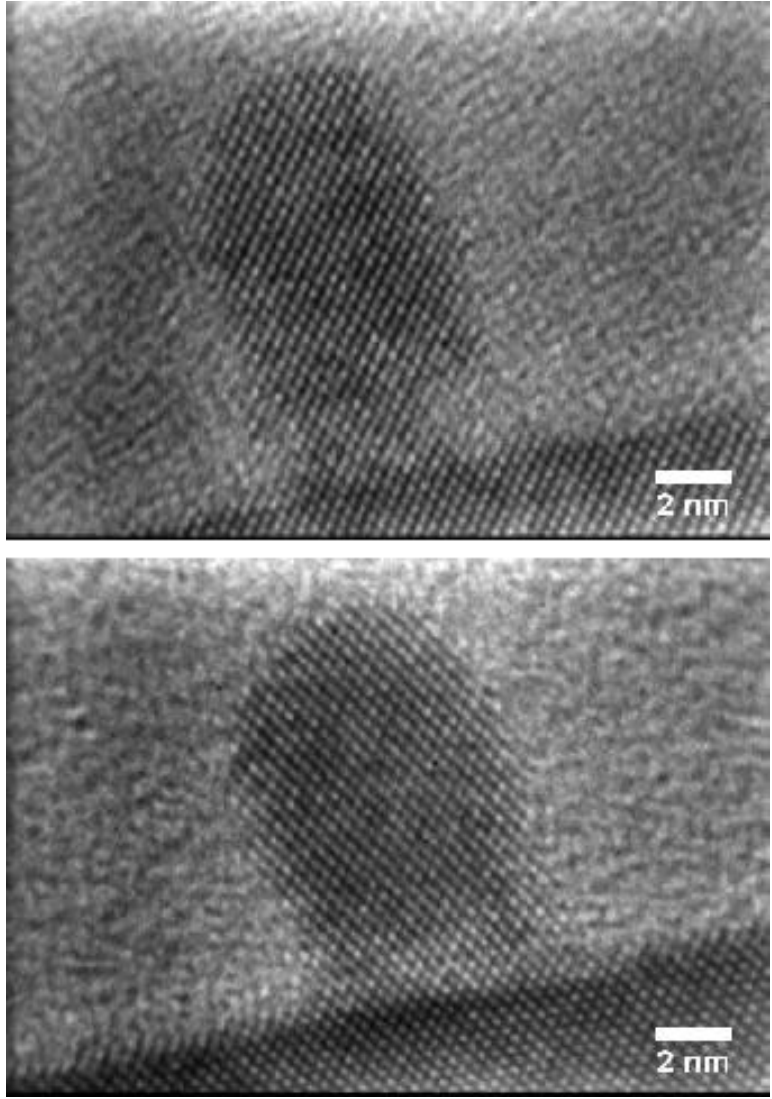

**Fig. 10.** HRXTEM micrographs obtained from a 2.7 BL Ge film grown at 550°C on an air-exposed clean Si(111)-(7×7) surface. Two typical vertically elongated epitaxial Ge islands are shown. These particular islands have an aspect ratio (diameter : height) of ~1:2.

# References

**1.** M. Zacharias and P. M. Fauchet, *Appl. Phys. Lett.* 71, 380 **(1997)**.

**2.** C. Bostedt, T.van Buuren, J. M. Plitzko, T. Moller and L. J. Terminello, *J. Phys.: Condens. Matter* 15, 1017 **(2003)**.

**3.** Y. Nakamura, K.Watanabe, Yo Fukuzawa and M. Ichikawa, *Appl. Phys. Lett.* 87, 133119 **(2005)**.

**4.** Y. M. Niquet, G. Allan, C. Delerue and M. Lannoo, *Appl. Phys. Lett.* 77, 1182 **(2000)**.

**5.** C. Bostedt, T.van Buuren, T. M. Willey, N. Franco, L. J. Terminello, C. Heske, T. Moller, *Appl. Phys. Lett.* 84, 4056 **(2004)**.

**6.** A. A. Shklyaev and M. Ichikawa, *Surf. Sci.* 514, 19 **(2002)**.

**7.** A. Konchenko, Y. Nakayama, I. Matsuda, S. Hasegawa, Y. Nakamura and M. Ichikawa, *Phys. Rev. B* 73, 113311 **(2006)**.

**8.** Bert Voigtlander, *Surf. Sci. Rep.* 43, 127 **(2001)**.

**9.** Y. Liou, P. W. Su and Y. L. Shen, *Appl. Phys. Lett.* 90, 182508 **(2007)**.

**10.** Y. Liou, M. S. Lee and K. L. You, *Appl. Phys. Lett.* 91, 082505 **(2007)**.

**11.** G. Medeiros-Ribeiro, A. M. Bratkovski, T. I. Kamins, D. A. A. Ohlberg and R. S. Williams, *Science* 279, 353 **(1998)**.

**12.** T. I. Kamins, G. Medeiros-Ribeiro, D. A. A. Ohlberg and R. S. Williams, *J. Appl. Phys.* 85, 1159 **(1999)**.

**13.** T. I. Kamins, E. C. Carr, R. S. Williams and S. J. Rosner, *J. Appl. Phys.* 81, 211 **(1997)**.

**14.** D. K. Goswami, B. Satpati, P. V. Satyam and B. N. Dev, *Curr. Sci.* 84, 903 **(2003)**.

**15.** Shinji Takeoka, *Phys. Rev. B* 58, 7921 **(1998)**.

**16.** A. A. Shklyaev, M. Shibata and M. Ichikawa, *Phys. Rev. B* 62, 1540 **(2000)**.

**17.** Y. Nakayama, I. Matsuda, S. Hasegawa and M. Ichikawa, *Appl. Phys. Lett.* 88, 253102 **(2006)**.

**18.** Y. Nakayama, S. Yamazaki, H. Okino, T. Hirahara, I. Matsuda, S. Hasegawa, M. Ichikawa, *Appl. Phys. Lett.* 91, 123104 **(2007)**.

**19.** Amal K. Das, J. Kamila, B. N. Dev, B. Sundaravel and G. Kuri, *Appl. Phys. Lett.* 77, 951 **(2000)**.

**20.** H. -C. Chung, W. -H. Chu and C. -P. Liu, *Appl. Phys. Lett.* 89, 082105 **(2006)**.

**21.** K. Sekar, P. V. Satyam, G. Kuri, D. P. Mahapatra and B. N. Dev, *Nucl. Instr. Meth. B* 71, 308 **(1992)**.

**22.** K. Sekar, P. V. Satyam, G. Kuri, D. P. Mahapatra and B. N. Dev, *Nucl. Instr. Meth. B* 73, 63 **(1993)**.

**23.** K. Sekar, P. V. Satyam, B. Sundaravel, G. Kuri, D. P. Mahapatra and B. N. Dev, *Phys. Rev. B* 51, 14330 **(1995)**.

**24.** K. Sekar, P. V. Satyam, B. Sundaravel, G. Kuri, D. P. Mahapatra and B. N. Dev, *Surf. Sci.* 339, 96 **(1995)**.

**25.** R. S. Becker, J. A. Golovchenko, G. S. Higashi and B. S. Swartzentruber, *Phys. Rev. Lett.* 57, 1020 **(1986)**.

**26.** M. J. Hadley and Steven P. Tear, *Surf. Sci. Lett.* 247, L221 **(1991)**.

**27.** Q. J. Gu, Z. L. Ma, N. Liu, X. Ge, W. B. Zhao, Z. Q. Xue, S. J. Pang and Z. Y. Hua, *Surf. Sci.* 327, 241 **(1995)**.

**28.** J. E. Demuth, R. J. Hamers, R. M. Tromp and M. E. Welland, *J. Vac. Sci. Technol. A* 3, 1320 **(1986)**.

**29.** H. Q. Yang, J. N. Gao, Y. F. Zhao, Z. Q. Xue and S. J. Pang, *Surf. Sci.* 406, 229 **(1998)**.

**30.** M. Hoshino, Y. Shigeta, K. Ogawa and Y. Homma, *Surf. Sci.* 365, 29 **(1996)**.

**31.** H. Tanaka, M. Udagawa, M. Itoh, T. Uchiyama, Y. Watanabe, T. Yokotsuka and I. Sumita, *Ultramicroscopy* 42-44, 864 **(1992)**.

**32.** K. Sekar, G. Kuri, D. P. Mahapatra, B. N. Dev, J. V. Ramana, Sanjiv Kumar and V. S. Raju, *Surf. Sci.* 302, 25 **(1994)**.

**33.** T. Maeda, A. Kurokawa, K. Sakamato, A. Ando, H. Itoh and S. Ichimura, *J. Vac. Sci. Technol. B* 19, 589 **(2001)**.

**34.** H. Iback, K. Horn, R. Dorn and H. Luth, *Surf. Sci.* 38, 433 **(1973)**.

**35.** K. Fujita, H. Watanabe and M. Ichikawa, *J. Appl. Phys.* 83, 3638 **(1998)**.

**36.** Ph. Avouris and R. Wolkow, *Phys. Rev. B* 39, 5091 **(1989)**.

**37.** L. Gavioli, K. R. Kimberlin, M. C. Tringides, J. F. Wendelken and Z. Zhang, *Phys. Rev. Lett.* 82, 129 **(1999)**.

**38.** D. K. Goswami, K. Bhattacharjee, B. Satpati, S. Roy, P. V. Satyam and B. N. Dev, *Surf. Sci.* 601, 603 **(2007)**.

**39.** T. Ohta, A. Klust, J. A. Adams, Q. Yu, M. A. Olmstead and F. S. Ohuchi, *Phys. Rev. B* 69, 125322 **(2004)**.

**40.** R. M. Stroud, A. T. Hanbicki, Y. D. Park, G. Kioseoglou, A. G. Petukhov and B. T. Jonker, *Phys. Rev. Lett.* 89, 166602 **(2002)**.

**41.** U. Kohler, J. E. Demuth and R. J. Hamers, *J. Vac. Sci. Technol. A* 7, 2860 **(1989)**.

**42.** A. A. Shklyaev and M. Ichikawa, in *Handbook of Semiconductor Nanostructures and Nanodevices*, edited A. A. Balandin and K. L. Wang, California, vol. 1, p. 337 **(2006)**.

**43.** Anupam Roy, K. Bhattacharjee, Trilochan Bagarti, K. Kundu and B. N. Dev (*to be published*)

**44.** Trilochan Bagarti, Anupam Roy, Kalyan Kundu and B. N. Dev, *Proceedings of Third National Conference on Mathematical Techniques: Emerging Paradigms for Electronics and IT Industries*. New Delhi, p. TS-2.1.1, January 30-31 **(2010)**.

**45.** A. Roy, K. Bhattacharjee, J. K. Dash and B. N. Dev, *Appl. Surf. Sci.* 256, 361 **(2009)**.

**46.** Anupam Roy, K. Bhattacharjee, J. Ghatak and B. N. Dev, (*to be published*); arXiv:1009.2633v1

**47.** Anupam Roy, J. K. Dash, A. Rath and B. N. Dev (*to be published*)

**48.** N. Shimizu, Y. Tanishiro, K. Kobayashi, K. Takayanagi and K. Yagi, *Ultramicroscopy* 18, 453 **(1985)**.

**49.** A. A. Shklyaev, M. Aono and T. Suzuki, *Surf. Sci.* 423, 61 **(1999)**.

**50.** A. V. Kolobov, A. A. Shklyaev, H. Oyanagi, P. Fons, S. Yamasaki and M. Ichikawa, *Appl. Phys. Lett.* 78, 2563 **(2001)**.